\documentstyle[11pt]{article}
\begin{document}
\begin{flushright}
\vspace{-3.0ex}
    {ADP-00-25/T408} \\
\vspace{-2.0mm}
\vspace{3.0ex}
\end{flushright}
\centerline{\bf\large ELECTROMAGNETIC RADIATION OF BARYONS}

\vspace{0.4cm}

\centerline{\bf\large CONTAINING TWO HEAVY QUARKS}

\vspace{1cm}

\centerline{
Wu-Sheng Dai$^{1,2,3}$, Xin-Heng Guo$^{4}$, Hong-Ying Jin$^{5}$ and 
Xue-Qian Li$^{1,2}$}

\vspace{1cm}

\begin{center}

{\small

1. CCAST (World Laboratory), P.O.8730, Beijing 100080, China\\

2. Department of Physics, Nankai University, Tianjin 300071, China\\

3. Department of Applied Physics, Tianjin University, Tianjin 300072, China\\

4. Department of Physics and Mathematical Physics, and Special
Research Center for the Subatomic Structure of Matter, University
of Adelaide, SA 5005, Australia\\

5. Institute of High Energy Physics, Beijing 100039, China}

\end{center}

\vspace{0.5cm}

\begin{center}
\begin{minipage}{11cm}
{\small
\noindent {\bf Abstract}

The two heavy quarks in a baryon which contains two heavy quarks and a light one,
can constitute a scalar or axial vector diquark.
We study electromagnetic radiations of such baryons,
(i) $\Xi_{(bc)_1}\rightarrow\Xi_{(bc)_0}+\gamma$, (ii)
$\Xi_{(bc)_1}^*\rightarrow\Xi_{(bc)_0}+\gamma$,
(iii) $\Xi_{(bc)_0}^{**}(1/2, l=1)\rightarrow\Xi_{(bc)_0}+\gamma$,
(iv) $\Xi_{(bc)_0}^{**}(3/2, l=1)\rightarrow\Xi_{(bc)_0}+\gamma$ and
(v) $\Xi_{(bc)_0}^{**}(3/2, l=2)\rightarrow\Xi_{(bc)_0}+\gamma$,
where $\Xi_{(bc)_{0(1)}}, \Xi^*_{(bc)_1}$ are
S-wave bound states of a heavy scalar or axial vector 
diquark and a light quark, and
$\Xi_{(bc)_0}^{**}(l\geq 1)$ are P- or D-wave bound states of a heavy scalar
diquark and a light
quark. Analysis indicates that these processes
can be attributed into two categories and the physical mechanisms
which are responsible for them are completely distinct.
Measurements can provide a good judgment
for the diquark structure and better understanding of the physical picture.}
\end{minipage}
\end{center}

\vspace{0.5cm}

\baselineskip 22pt
\noindent{PACS numbers: 12.39.Hg, 11.10.St, 13.40.Hq, 13.30.-a}

\newpage

\noindent{\bf I. Introduction}\\

Lack of an effective way to properly handle non-perturbative QCD effects
becomes a more and more intriguing problem when one needs to extract
information from data. In other words, the hadronic matrix elements cannot be
reliably estimated in the present theoretical framework. Thanks to the
heavy quark effective theory (HQET) \cite{Isgur}, an extra symmetry $SU(2)_s
\otimes SU(2)_f$ greatly simplifies the picture in the heavy flavor involved
processes. Developments in this field enable us to
more accurately evaluate hadronic transition matrix elements since the number
of form factors is reduced in the heavy quark limit \cite{Dai}.
%as only one unknown
%form factor, the Isgur-Wise function $\xi(v\cdot v')$ remains. On other side,
%in the framework of the HQET, the authors of ref.\cite{Dai} indicated that
%there are many independent form factors in baryonic transition matrix
%elements even at order of $O((1/M_Q)^0)$, so that  an accurate theoretical
%computation of such processes involving baryons become very difficult unless
%one can employ some models to simplify the picture.

As many authors suggested, there may exist the diquark structure in baryons
\cite{Ebert}. If it is the real physics, or at least a good approximation,
we only need to deal with two-body problems instead of three-body one.
Cosequently, the number of
independent form factors can be remarkably reduced.
Especially, when the baryons contain two heavy quarks, it is reasonable to
assume that the two heavy quarks constitute a color-anti-triplet
boson-like diquark of spin 0 or 1 \cite{Falk}. Based on this picture Savage and
Wise studied the spectrum of baryons with two heavy quarks \cite{Sav} and
in the potential model, the spectra have been evaluated \cite{Ebert1}.

Althoug the diquark structure is very likely, the small color-anti-triplet
system is not point-like in general. Consequently, we should replace the vertex
gained from
any fundamental theory such as the Standard Model by an effective vertex.
A (or a few) reasonable form factor(s)
will be involved in the effective vertex for compensating the
non-point-like spatial dispersion of the diquark. The form factor(s)
can be derived in many
ways, and one of them is the Bethe-Salpeter
(B-S) equation. With the effective vertex,
we estimated the production and weak decay rates of such baryons \cite{Guo}
in our previous work based on the superflavor symmetry \cite{Georgi1,Carone}.
%In this scenario, one can reduce the unknown
%hadronization factors which represents the non-perturbative QCD effects to
%the unique Isgur-Wise function.

To further investigate the diquark structure and the governing mechanisms
inside the diquark, we will study electromagnetic radiations of baryons with
two heavy quarks in the present work. Since
such processes are cleaner, we may expect to gain more exact knowledge from the
data. In fact, similar electromagnetic radiation processes for baryons
containing only one heavy quark have been discussed in literature
recently \cite{Ivanov}.

At the tree level, the $\gamma-$emission is a pure electromagnetic process.
In this work we study two cases which in fact are determined by completely
different mechanisms.
First, we consider (i) $\Xi_{(bc)_1}\rightarrow\Xi_{(bc)_0}+\gamma$ and
(ii) $\Xi_{(bc)_1}^*\rightarrow\Xi_{(bc)_0}+\gamma$, where $\Xi_{(bc)_1}$ and 
$\Xi^*_{(bc)_1}$ are spin 1/2 and 3/2 baryons (respectively)
which consist of a heavy axial
vector diquark and a light
quark in S-wave, and $\Xi_{(bc)_0}$ is a spin-1/2 baryon which consists of a
heavy scalar diquark
and a light quark. Then we study (iii) $\Xi_{(bc)_0}^{**}(1/2,l=1)
\rightarrow\Xi_{(bc)_0}+\gamma$, (iv)  $\Xi_{(bc)_0}^{**}(3/2,l=1)
\rightarrow\Xi_{(bc)_0}+\gamma$ and (v) $\Xi_{(bc)_0}^{**}(3/2,l=2)
\rightarrow\Xi_{(bc)_0}+\gamma$ where
$\Xi_{(bc)_0}^{**}(s,l\geq 1)$ are spin 1/2 ($s=1/2$) and 3/2 ($s=3/2$)
baryons (respectively) composed
of a heavy scalar diquark and a light quark in higher angular momentum states.
It is noted that we study the $(bc)_{1(0)}$ diquark because only
$(bc)$ can constitute either spin 1 or 0 states with even parity (i.e., 
the orbital angular momentum between $Q$ and $Q'$ is set to be 0 in our
discussion).

In the reactions (i) $\Xi_{(bc)_1}\rightarrow\Xi_{(bc)_0}+\gamma$ and
(ii) $\Xi_{(bc)_1}^*\rightarrow\Xi_{(bc)_0}+\gamma$,
the axial vector
$(bc)_1$ transits into a scalar $(bc)_0$ by emitting a photon;
whereas in the radiations (iii) $\Xi_{(bc)_0}^{**}(1/2,l=1)
\rightarrow\Xi_{(bc)_0}+\gamma$, (iv)  $\Xi_{(bc)_0}^{**}(3/2,l=1)
\rightarrow\Xi_{(bc)_0}+\gamma$, and (v) $\Xi_{(bc)_0}^{**}(3/2,l=2)
\rightarrow\Xi_{(bc)_0}+\gamma$,
the diquark $(bc)_0$ remains in the
spin-0 state, and the photon is radiated from the light quark hand. The later
three reactions are in analog to the radiation of atom where electron transits
from a higher (angular and/or radial) exited state  into
a lower one and emits a photon.
In our case, the light quark of $\Xi^{**}_{(bc)_0}(s,l\geq 1)$ in an
angular-momentum
excited state transits into the ground state ($l=0$) $\Xi_{(bc)_0}$ and emits a
photon. Analysis indicates that
the possibility of radiating a photon from
the spin-0 heavy diquark is very small, exactly as in
the case of atoms.

Of course, in general, there may be processes like $\Xi_{(bc)_1}^*(3/2)
\rightarrow\Xi_{(bc)_1}(1/2)+\gamma$. However, since
the spin interaction between gluon and heavy diquarks decouples in the heavy
quark limit, the mass splitting between
$\Xi^*_{(bc)_1}$ and $\Xi_{(bc)_1}$  is 0. Consequently, in the heavy
quark limit, the radiative
transition between these two states is forbidden by the null phase space.
So we do not discuss such processes in this work.

In the next section, we present our formulation for the two different
radiation mechanisms
and in the third section, we give the numerical results. The last section
is devoted to discussion and conclusion and finally in the appendix, we give
all the concerned expressions which are omitted in the context.\\

\noindent{\bf II. Formulation}\\

In this section, we discuss the two different mechanisms respectively.

(a) Radiation from the heavy diquark hand.

As discussed in the introduction, for the radiation processes
$\Xi_{(bc)_1}\rightarrow \Xi_{(bc)_0}+\gamma$ and
$\Xi^{*}_{(bc)_1}
\rightarrow \Xi_{(bc)_0}+\gamma$, the axial vector diquark transits into a scalar
diquark by emitting a photon and the light quark remains as a spectator. In
this case, all the non-perturbative effects can be attributed into a form
factor at the leading order of expansion with respect to the heavy quark
mass. To evaluate the transition matrix elements, we
employ the superflavor symmetry \cite{Georgi1,Carone}, which is applicable
to this situation.

At the effective vertex $AS\gamma$, where $A$ and $S$ denote axial vector
and scalar diquarks 
(respectively) and  $\gamma$ is the emitted photon, a form factor
can be derived
in terms of the B-S equation \cite{Guo}. The transition amplitude can be
written as
\begin{equation}
T=\epsilon^*_{\alpha}\langle J^{\alpha} \rangle,
\end{equation}
where $\epsilon^*_{\alpha}$ is the polarization vector of the axial vector
diquark, and $J^{\alpha}$ is the effective current at the quark level and
$\langle J^{\alpha}\rangle$ is the corresponding transition amplitude.

For $\Xi_{(bc)_1}\rightarrow \Xi_{(bc)_0}+\gamma$,
\begin{equation}
\langle J^{\alpha}\rangle= \langle \Xi_{(bc)_0}(v')|J^{\alpha}
|\Xi_{(bc)_1}(v)\rangle
= \xi(v'\cdot v)if\epsilon^{\alpha\delta\rho\sigma}v_{\rho}v_{'\sigma}
\bar u'(v')\gamma_5\gamma_{\delta}u(v),
\end{equation}
and for $\Xi^*_{(bc)_1}\rightarrow \Xi_{(bc)_0}+\gamma$,
\begin{equation}
\langle J^{\alpha}\rangle = \langle \Xi_{(bc)_0}(v')|J^{\alpha}
|\Xi^*_{(bc)_1}(v)\rangle
= \xi(v'\cdot v)if\epsilon^{\alpha\delta\rho\sigma}v_{\rho}v_{'\sigma}
\bar u'(v')u_{\delta}(v),
\end{equation}
where $\xi(v\cdot v')$ is the Isgur-Wise function, $v$ and $v'$ are the
four-velocities of the parent and daughter baryons, respectively,
$u$ is the four-component
spinor for the parent or produced baryon $\Xi_{(bc)_{1(0)}}$,
and $u_{\delta}$ is the
Rarita-Schwinger spinor-vector corresponding to $\Xi_{(bc)_1}^{*}$ with
spin 3/2. The form factor is evaluated in the B-S equation approach
and all the details were given in our previous work \cite{Guo}.

Taking the amplitude square, we have:
for $\Xi_{(bc)_1}\rightarrow \Xi_{(bc)_0}+\gamma$,
\begin{equation}
\label{1/2}
{1\over 2}\sum_{all \; spins}|T|^2={e^2\over 36}|\xi(v\cdot v')|^2Tr[
C^{\rho\sigma}C^{\rho'\sigma'}\epsilon_{\alpha\delta\rho\sigma}\epsilon_{
\alpha'\delta'\rho'\sigma'}\bar u'\gamma_5\gamma^{\delta}u
\bar u\gamma^{\delta'}\gamma_5u']\sum_{\lambda}
\epsilon^{\alpha}_{(\lambda)}\epsilon^{*\alpha'}_{(\lambda)},
\end{equation}
and for $\Xi^*_{(bc)_1}\rightarrow \Xi_{(bc)_0}+\gamma$,
\begin{equation}
\label{3/2}
{1\over 4}\sum_{all \; spins}|T|^2={e^2\over 36}|\xi(v\cdot v')|^2Tr[
C^{\rho\sigma}C^{\rho'\sigma'}\epsilon_{\alpha\delta\rho\sigma}\epsilon_{
\alpha'\delta'\rho'\sigma'}\bar u'u^{\delta}\bar u^{\delta'}u']\sum_{\lambda}
\epsilon^{\alpha}_{(\lambda)}\epsilon^{*\alpha'}_{(\lambda)},
\end{equation}
where
\begin{equation}
C^{\rho\sigma}=fv^{\rho}v^{'\sigma},
\end{equation}
and numerically
$$f\sim 1.$$
In our case, the photon emitted from the heavy diquark only carries very
small momentum and energy, thus $v\cdot v'$ would be very close to unity, so
$$\xi(v\cdot v')\approx 1.$$

Then we can easily obtain the widths of these radiative decay processes as
\begin{equation}
\Gamma={1\over 2M}\int {d^3p\over (2\pi)^3}{1\over 2E}{d^3k\over (2\pi)^3}{1\over 2\omega}
(2\pi)^4\delta^4(P-p-k){1\over 2s+1}\sum_{all\; spins}|T|^2,
\end{equation}
where $P,p,k$ are the four-momenta of the initial, final baryons and emitted
photon, respectively, and $M$ is mass of the initial baryon.
Because it is a two-body final state case, the integration is
very easy to be carried out.

(b) Radiation from the light quark hand.

In this case, $\Xi^{**}_{(bc)_0}(s,l\geq 1)$ are composed of a scalar diquark and a light
quark in a higher angular momentum state ($l\geq 1$), thus the radiation
is realized via a process
that the light quark transits from a higher angular momentum state into
the ground state ($l=0$) via emitting a photon. This process is in analog to
the photon radiation of atoms where the electron jumps from an excited state
(radial or $l\geq 1$) into
the ground state via emitting a photon.

In these processes, the heavy diquark acts as a spectator. Since the reaction
happens on the light flavor side, HQET is not applicable in this case.
Instead, we use the B-S equation to calculate the transition matrix
elements. For consistency, the wavefunctions of
$\Xi^{**}_{(bc)_0}(1/2(3/2),l=1,2)$
and
$\Xi_{(bc)_0}(1/2,l=0)$ are also obtained in terms of the B-S equation. The
wavefunctions are given in the following:
\begin{eqnarray}
\label{1/20}
\kappa_P^{(1/2,0)}(p) &=& (\phi^{(10)}_1(p)+\phi^{(10)}_2(p)\rlap /p_t)u(P),
\;\;\;\;(s={1\over 2},l=0)\\
\label{1/21}
\kappa_P^{(1/2,1)}(p) &=& (\phi^{(11)}_1(p)+\phi^{(11)}_2(p)\rlap /p_t)
\gamma_5u(P),\;\;\;\; (s={1\over 2},l=1)\\
\label{3/21}
\kappa_P^{(3/2,1)}(p) &=& (\phi^{(31)}_1(p)+
\phi^{(31)}_2(p)\rlap /p_t)p_{t\mu}u^{\mu}(P), \;\;\;\;(s={3/2},l=1)\\
\label{3/22}
\kappa_P^{(3/2,2)}(p) &=& [\phi^{(32)}_1(p)+\phi^{(32)}_2(p)\rlap /p_t]
\gamma_5p_{t\mu}u^{\mu}(P),
\;\;\;\;(s={3\over 2},l=2),
\end{eqnarray}
where $u(P)$ is the spinor for the baryon of spin-1/2 and
$u^{\mu}(P)$ is the Rarita-Schwinger spinor-vector. Here we use the transverse
momentum $p_t$ which is defined as
\begin{equation}
p_t^{\mu}=p^{\mu}-p_lv^{\mu},
\end{equation}
and $v^{\mu}$ is the four-velocity of the concerned baryon, $p_l\equiv p\cdot v$ is
the longitudinal momentum.

The  vertex $\bar qq\gamma$ is the typical QED coupling. Taking the loop
integration with the obtained B-S wavefunctions we can have the transition
amplitude square as the following.

For $\Xi^{**}_{(bc)_0}(1/2,l=1)\rightarrow \Xi_{(bc)_0}+\gamma$,
\begin{equation}
{1\over 2}\sum_{all\; spins}|T|^2={e^2\over 2}\sum_{all\; spins}
|\bar u(v')G^{\mu}u(v)\epsilon_{\mu}^{(\lambda)*}|^2,
\end{equation}
where
\begin{equation}
\label{Gi}
G^{\mu}\equiv G_1\gamma^{\mu}\gamma_5+G_2\gamma^{\mu}\rlap /v'_t\gamma_5
+G_3\rlap /v'_t\gamma^{\mu}\gamma_5+G_4(-2\gamma^{\mu}\gamma_5+\rlap /v
\gamma^{\mu}\gamma_5)+G_5
\rlap /v'_t\gamma^{\mu}\rlap /v'_t\gamma_5.
\end{equation}

For $\Xi^{**}_{(bc)_0}(3/2,l=1)\rightarrow \Xi_{(bc)_0}+\gamma$,
\begin{equation}
{1\over 4}\sum_{all\; spins}|T|^2={e^2\over 4}\sum_{all\; spins}|\bar u(v')
H_{\nu}^{\mu}u^{\nu}(v)
\epsilon_{\mu}^{(\lambda)*}|^2,
\end{equation}
where
\begin{equation}
\label{Hi}
H^{\mu}_{\nu}\equiv H_1\gamma^{\mu}v'_{\nu}+H_3\gamma^{\mu}\rlap /v'_tv'_{\nu}
+2H_4g^{\mu}_{\nu}+H_5\rlap /v'_t\gamma^{\mu}v'_{\nu}
+H_7\rlap /v_t\gamma^{\mu}\rlap /v'_tv'_{\nu}.
\end{equation}

For $ \Xi^{**}_{(bc)_0}(3/2,l=2)\rightarrow \Xi_{(bc)_0}+\gamma$,
\begin{equation}
{1\over 4}\sum_{all\; spins}|T|^2={e^2\over 4}\sum_{all\; spins}
|\bar u(v')F_{\nu}^{\mu}\gamma_5u^{\nu}(v)
\epsilon_{\mu}^{(\lambda)*}|^2,
\end{equation}
where
\begin{equation}
\label{Fi}
F_{\nu}^{\mu}=F_1\gamma^{\mu}v'_{\nu}+F_3\gamma^{\mu}\rlap /v'_tv'_{\nu}
+2F_4g_{\nu}^{\mu}+F_5\rlap /v'_t\gamma^{\mu}v'_{\nu}+
F_7\rlap /v_t\gamma^{\mu}\rlap /v'_tv_{\nu}'.
\end{equation}

All the coefficients $G_i$, $H_i$ and $F_i$ in eqs.(\ref{Gi},\ref{Hi},\ref{Fi})
are related to the B-S integrals and we give their explicit
expressions in the appendix. The derivation in terms of the B-S equation is
very tedious but standard.

The partial width is obtained in the same way as in II(a).\\

\noindent{\bf III. The numerical results}

(a) Radiation from the heavy diquark hand.

%(i) For $\Xi^{*}_{(bc)}(3/2)\rightarrow \Xi_{(bc)}(1/2)+\gamma$.

Since there are no data for the masses of baryons containing two heavy quarks
yet, we have to take the theoretically estimated values which are given in
literatures. Here we use the results given by Ebert et al. \cite{Ebert1} as
$M_{\Xi_{(bc)}^*}=7.02$ GeV and $M_{\Xi_{(bc)}}=6.95$ GeV.
We have
$$\Gamma(\Xi_{(bc)_1}\rightarrow \Xi_{(bc)_0}+\gamma)
\sim 2.75\times 10^{-9}
\; {\rm GeV},$$
$$\Gamma(\Xi^{*}_{(bc)_1}\rightarrow \Xi_{(bc)_0}+\gamma)\sim
7.48\times 10^{-9}\;{\rm GeV}.$$
Namely, the widths are in order of eV's.

(b) Radiation from the light quark hand.

For consistency, we have also obtained the binding energies of the concerned
baryons in terms of the B-S equation. We have
$$E_{\Xi^{**}_{(bc)_0}(3/2,l=2)}=1.39\;{\rm GeV},\;\;
E_{\Xi^{**}_{(bc)_0} (3/2,l=1)}=0.69\;{\rm GeV},$$
$$E_{\Xi^{**}_{(bc)_0}(1/2,l=1)}=0.66\;{\rm GeV},\;\;
E_{\Xi_{(bc)_0}(1/2,l=0)}=0.026\;{\rm GeV}.$$

In this framework, we have
$$M_{\Xi_{(bc)_0}^{**}(s,l)}=m_1+m_2+E_{\Xi_{(bc)_0}^{**}(s,l)},$$
$$M_{\Xi_{(bc)_0}}=m_1+m_2+E_{\Xi_{(bc)_0}},$$
where $m_1$ and $m_2$ are  the masses of the light quark and the heavy
scalar diquark, respectively, $E$ is the binding energy.
To evaluate the binding energies,
we take the simplest potential form which contains only the Coulomb and linear
confinement pieces as the B-S kernel \cite{Guo}.

Numerically, we take
$$ m_1=0.33\; {\rm GeV} \;({\rm for\; u-\; and \; d-quark}),\; 0.5 \; 
{\rm GeV}\;({\rm for \; s-quark});\;\;\; m_2=6.52\; {\rm GeV},$$
as inputs \cite{Ebert}.

We use these values in the numerical evaluations and obtain:
$$\Gamma(\Xi^{**}_{(bc)_0}(1/2,l=1)\rightarrow \Xi_{(bc)_0}(1/2)+
\gamma)\sim 1.5\times 10^{-4} \;{\rm GeV},$$
$$\Gamma(\Xi^{**}_{(bc)_0}(3/2,l=1)\rightarrow \Xi_{(bc)_0}(1/2)+
\gamma)\sim 3.7\times 10^{-5} \;{\rm GeV},$$
$$\Gamma(\Xi^{**}_{(bc)_0}(3/2,l=2)\rightarrow \Xi_{(bc)_0}(1/2)+
\gamma)\sim 6.2\times 10^{-4} \;{\rm GeV}.$$

As discussed above, these partial widths are evaluated in terms of
the B-S equation.
Indeed these reactions  are governed by a mechanism different from that in (a),
and the methods we use for evaluating the widths are distinct.

In this subsection, we obtain the masses of
$\Xi_{(bc)_0}^{**}(1/2(3/2),l\geq 1)$ and $\Xi_{(bc)_0}(1/2)$ and
the transition matrix element
$\langle \Xi_{(bc)_0}(1/2)|J_{\mu}|\Xi_{(bc)_0}^{**}(1/2(3/2),l\geq 1)
\rangle$ in the same framework, i.e.
the B-S equation. In fact, there is no any substantial difference from the
values we take in subsection (a) for $\Xi_{(bc)_1}\rightarrow \Xi_{(bc)_0}
+\gamma$ and $\Xi^{*}_{(bc)_1}\rightarrow \Xi_{(bc)_0}+\gamma$.

It is noted that the mass difference between the angular-momentum excited
state $\Xi_{(bc)_0}^{**}(3/2,l\geq 1)$ and the ground S-state $\Xi_{(bc)_0}$ is
about 0.6$\sim$1.4 GeV. It is much larger than that between $\Xi_{(bc)_1}^{*}$
and $\Xi_{(bc)_0}$ (0.07 GeV). It is easy to understand: the former
one is due to the orbital angular momentum excitation and the later one is due
to an energy splitting between axial vector and scalar diquarks, which is
caused by the spin-spin interaction. Therefore for $\Xi^{**}_{(bc)_0}(s,l\geq 1)
\rightarrow\Xi_{(bc)_0}+\gamma$
the threshold effects are not
obvious and the widths are about 4 orders of magnitude larger than
$\Gamma(\Xi_{(bc)_1}(\Xi^{*}_{(bc)_1})\rightarrow \Xi_{(bc)_0}(1/2)+\gamma)$. In other words,
the remarkable width difference for the two processes is due to the threshold
effects while the matrix elements for both reactions are of the same order
of magnitude.\\

\noindent{\bf IV. Conclusion and discussion}\\

HQET is proved to be effective in many processes where heavy flavors are
involved. In most cases, the light flavors in the hadrons just behave as
spectators for the reactions and these degrees of freedom manifest in
the hadronization processes, and therefore determine
the form factors such as the
Isgur-Wise function. However, in some cases, the light flavors may
participate in reactions and sometimes can play a crucial  role. As we
know, when the quark level final state interaction is involved, the
W-annihilation and especially the Pauli Interference can be very
important in the inclusive B meson decays \cite{Bigi}, then the
contribution from the light
flavor could be as important as that from the heavy one.

In this work, we choose two different kinds of processes where the heavy
and light flavors are active respectively. $\Xi_{(bc)_1}$ and
$\Xi_{(bc)_1}^{*}$
consist of an axial vector diquark and a light quark. When they transit into
$\Xi_{(bc)_0}$ by radiating a photon, the axial vector diquark
turns into a scalar one, and the light quark serves as a spectator
in this process.
On the contrary, $\Xi_{(bc)_0}^{**}(1/2(3/2),l\geq 1)$ consists of a scalar
heavy diquark, and a light quark at
angular-momentum excited states ($l=1,2$ in this work). Thus when it
transits into
$\Xi_{(bc)_0}$, the heavy diquark stands as a spectator and the light
quark jumps from a higher-excited state into the ground state while radiating
a photon. For the former one, HQET definitely applies and by the
superflavor symmetry, we can expect to obtain a more accurate result of the
decay width. Once the doubly heavy baryon masses are
measured,  we can immediately have
the final numbers with our formula for the partial width.
As long as HQET works, the result should be close to
data. Of course, there is also an uncertain factor, it is the form factor at
the effective vertex of $SA\gamma$. We obtain it in terms of the B-S equation,
where the potential kernel would bring up some uncertainty. However, in this
case, the diquark is composed of two heavy quarks, so the non-relativistic
Cornell potential works well as understood.
Moreover, careful studies indicate that for so small recoil situation,
$(v\cdot v')\sim 1$, the form factor $f$ is close to 1.
%This conclusion does not depend on the details
%of how to evaluate the form factor $f$. 
Therefore,
we can expect that the relative errors for the partial widths of
$\Xi_{(bc)_1}\rightarrow \Xi_{(bc)_0}+\gamma$ and
$\Xi^{*}_{(bc)_1}\rightarrow \Xi_{(bc)_0}+\gamma$ are quite small.
The widths are of order of eV's and similar to that for atomic radiation.
The smallness is easy to understand. Let us use
$\Xi^{*}_{(bc)_1}\rightarrow \Xi_{(bc)_0}+\gamma$ as an example. From
eq.(\ref{3/2}), we have
\begin{equation}
 {1\over 4}\sum_{all\; spins}|T|^2={4e^2\over 27}f^2M_{1/2} M'_{3/2}
 (v\cdot v'-1)(1+v'\cdot v)^2,
 \end{equation}
where $M_{1/2}$ and $M'_{3/2}$ are the masses of $\Xi_{(bc)}(1/2)$ and
$\Xi^*_{(bc)}(3/2)$
respectively. In this case, $v\cdot v'-1$ is close to zero and it is nothing but
the threshold effect. With this expression, we can easily obtain the partial width
 of this radiative decay as
 \begin{equation}
 \label{ga1}
\Gamma={\alpha\over 216}f^2{(M_{3/2}^{' 2}-M_{1/2}^2)^3\over M_{3/2}^{' 5}
M_{1/2}^2}
 (M'_{3/2}+M_{1/2})^2.
 \end{equation}
It is noted that the width is proportional to $(M'^{2}_{3/2}-M_{1/2})^3/
M'^{5}_{3/2}$, hence
for small difference between the masses of the parent and daughter baryons, the
 threshold effects are very obvious. One can
expect this threshold effects to strongly suppress the width.

As a matter of fact, these radiative decay processes where the heavy axial
vector 
diquark emits a photon and transits into a scalar one, are in analog to the
radiative decay $J/\psi\rightarrow\eta_c+\gamma$ whose partial width is
about 1.13 KeV \cite{Data}. But there are several suppression factors in
the doubly heavy baryon case.
First in $J/\psi$ $c $ and $\bar c$ reside in a color singlet, but in the
diquark, $b$ and $c$ quarks are in a color $\bar 3$ state, there should be a factor
1/8 suppression for the diquark transition. From
the formula (\ref{ga1}), one has
a factor $(M'^2_{3/2}-M^2_{1/2})/M'^{5}_{3/2}$,
so totally there could be a suppression of
about $5\times 10^{-3}$ compared to the $J/\psi$ radiative decay. The net
result is of eV order.

For $\Xi^{**}_{(bc)_0}(1/2(3/2),l\geq 1)\rightarrow \Xi_{(bc)_0}
+\gamma$,
HQET does not apply and we need to
employ the B-S equation
method to evaluate the transition matrix elements.
In the calculations, the B-S wavefunctions of the initial and final states are
needed. Since in such radiative decays, the recoil energy-momentum of the
final baryon is very small compared to the involved energy scales, 
we expect the theoretical predictions are quite reliable.
%the
%instantaneous approximation for solving the B$-$S equation would not bring up
%sizable errors and the results are relatively reliable.

The numerical results show that for $\Xi_{(bc)_1}\rightarrow
\Xi_{(bc)_0}+\gamma$ and $\Xi^{*}_{(bc)_1}\rightarrow
\Xi_{(bc)_0}+\gamma$, the partial width is in order of eV, and for
$\Xi^{**}_{(bc)_0}(3/2(1/2), l=1(2))\rightarrow \Xi_{(bc)_0}(1/2)+\gamma$,
it is of 10$\sim$100
KeV. The difference is due to the threshold effects.

Besides the study on the reaction mechanisms,  this work also concerns
testifying the diquark structure in baryons. It is believed that the two
heavy quarks inside a baryon can constitute a diquark of scalar or axial vector
which is a relatively stable physical subject \cite{Guo}.
Our calculations are based on
such a physical picture and the future experiments should test it.
Lack of data on baryons which consist of two heavy quarks so far makes
drawing a definite conclusion difficult. But it is possible that the data can
be accumulated in near future experiments. Once we have the data on the masses,
we can re-evaluate the numbers of decay widths easily. Then comparing the
calculated results with data, we can determine the validity of the diquark
structure  and the reaction mechanisms.
No doubt, the experiments for the electromagnetic  radiation are difficult, but
as suggested \cite{Savage}, the radiative decay may be measurable
soon, and the background in this case is clean. We believe that
the results can enrich our
knowledge on baryons, so is worth of careful investigations.\\

\noindent{\bf Acknowledgment:}

This work is supported in part by the National Natural Science Foundation
of China and the Australian Research Council.

\vspace{1cm}

\noindent{\bf Appendix}

Here we present the explicit expressions of the form factors $G_i$, $H_i$
and $F_i$ in eqs.(14,16,18).
\begin{equation}
G_i\equiv \int{dp_l\over 2\pi}g_i;\;\;\; H_i\equiv \int{dp_l\over 2\pi}h_i;
\;\;\;\; F_i\equiv \int{dp_l\over 2\pi}f_i.
\end{equation}
\begin{eqnarray}
g_1 &=& \int{d^3p_t\over (2\pi)^3}ac';  \nonumber \\
g_2 &=& {1\over \sqrt{(v\cdot v')^2-1}}
\int{d^3 p_t\over (2\pi)^3}ad'|\vec p_t|\cos\theta;\nonumber \\
g_3 &=& {1\over \sqrt{(v\cdot v')^2-1}}\int {d^3 p_t\over (2\pi)^3}
bc'|\vec p_t|\cos\theta;\nonumber \\
g_4 &=& {-1\over 2}\int {d^3 p_t\over (2\pi)^3}bd'|\vec p_t|^2(1-\cos^2\theta);
\nonumber\\
g_5 &=& {-1\over 2}{1\over (v\cdot v')^2-1}\int{d^3 p_t\over (2\pi)^3}bd'
|\vec p_t|^2(1-3\cos^2\theta);\nonumber\\
h_1 &=& {1\over \sqrt{(v\cdot v')^2-1}}\int{d^3p_t\over (2\pi)^3}ac''|\vec p_t|
\cos\theta; \nonumber \\
h_2 &=& {-1\over 2}\int{d^3 p_t\over (2\pi)^3}ad''|\vec p_t|^2(1-\cos^2\theta);
\nonumber \\
h_3 &=& {-1\over 2}{1\over (v\cdot v')^2-1}\int {d^3 p_t\over (2\pi)^3}
ad''|\vec p_t|^2(1-3\cos^2\theta);\nonumber \\
h_4 &=& {-1\over 2}\int {d^3 p_t\over (2\pi)^3}bc''|\vec p_t|^2(1-\cos^2\theta);
\nonumber\\
h_5 &=& {-1\over 2}{1\over (v\cdot v')^2-1}\int{d^3 p_t\over (2\pi)^3}bc''
|\vec p_t|^2(1-3\cos^2\theta);\nonumber\\
h_6 &=& {-1\over 2}\int{d^3 p_t\over (2\pi)^3}bd''(\lambda_2M_{(3/2,l=1)}+p_l)
|\vec p_t|^2(1-\cos^2\theta);\nonumber\\
h_7 &=& {-1\over 2}{1\over (v\cdot v')^2-1}\int {d^3 p_t\over (2\pi)^3}
bd''(\lambda_2M_{(3/2,l=1)}+p_l)|\vec p_t|^2(1-3\cos^2\theta);\nonumber\\
f_1 &=& {1\over \sqrt{(v\cdot v')^2-1}}
\int{d^3p_t\over (2\pi)^3}ac''|\vec p_t|
\cos\theta; \nonumber\\
f_2 &=& {-1\over 2}\int{d^3 p_t\over (2\pi)^3}ad|\vec p_t|^2(1-cos^2\theta);
\nonumber\\
f_3 &=& {-1\over 2}{1\over (v\cdot v')^2-1}\int {d^3 p_t\over (2\pi)^3}
ad|\vec p_t|^2(1-3\cos^2\theta);\nonumber\\
f_4 &=& {-1\over 2}\int {d^3 p_t\over (2\pi)^3}bc|\vec p_t|^2(1-\cos^2\theta);
\nonumber\\
f_5 &=& {-1\over 2}{1\over (v\cdot v')^2-1}\int{d^3 p_t\over (2\pi)^3}bc
|\vec p_t|^2(1-3\cos^2\theta);\nonumber\\
f_7 &=& {-1\over 2}{1\over (v\cdot v')^2-1}\int {d^3 p_t\over (2\pi)^3}
bd(\lambda_2M_{(3/2,l=2)}+p_l)|\vec p_t|^2(1-3\cos^2\theta),
\end{eqnarray}
where $\theta$ is the angle between $p_t$ and $v_t$,

\begin{eqnarray}
a &=& \left[{2\omega'_{p_t}(\omega'_{p_t}-m_1-E_{(1/2,l=0)})\over
E_{(1/2,l=0)}-p'_l)+i\epsilon}{2m_1+p'_l\over (p'_l+m_1)^2-\omega^{'2}_{p_t}+i\epsilon}
\right]\tilde\Phi^{(10)}_2;\nonumber\\
b &=&\left[{2\omega'_{p_t}(\omega'_{p_t}-m_1-E_{(1/2,l=0)})\over
E_{(1/2,l=0)}-p'_l)+i\epsilon}{1\over (p'_l+m_1)^2-\omega^{'2}_{p_t}+i\epsilon}
\right] \tilde\Phi^{(10)}_2;\nonumber\\
c &=& -i\left[{2\omega_{p_t}(\omega_{p_t}-m_1-E_{(3/2,l=2)})\over
E_{(3/2,l=2)}-p_l)+i\epsilon}{-p_l\over (p_l+m_1)^2-\omega^{2}_{p_t}+i\epsilon}
\right]\nonumber\\
&& \times (2m_2(p_l-E_{(3/2,l=2)}))\tilde\Phi^{(32)}_2;\nonumber\\
d &=& -i\left[{2\omega_{p_t}(\omega_{p_t}-m_1-E_{(3/2,l=2)})\over
(E_{(3/2,l=2)}-p_l)+i\epsilon}{1\over (p_l+m_1)^2-\omega^{2}_{p_t}+i\epsilon}
\right] \nonumber \\
&& \times (2m_2(p_l-E_{(3/2,l=2)}))\tilde\Phi^{(32)}_2;\nonumber\\
c' &=& -i\left[{2\omega_{p_t}(\omega_{p_t}-m_1-E_{(1/2,l=1)})\over
E_{(1/2,l=1)}-p_l)+i\epsilon}{-p_l\over (p_l+m_1)^2-\omega^{2}_{p_t}+i\epsilon}
\right] \nonumber\\
&& \times (2m_2(p_l-E_{(1/2,l=1)}))\tilde\Phi^{(11)}_2;\nonumber\\
d' &=& -i\left[{2\omega_{p_t}(\omega_{p_t}-m_1-E_{(1/2,l=1)})\over
E_{(1/2,l=1)}-p_l)+i\epsilon}{1\over (p_l+m_1)^2-\omega^{2}_{p_t}+i\epsilon}
\right] \nonumber\\
&& \times (2m_2(p_l-E_{(1/2,l=1)}))\tilde\Phi^{(11)}_2;\nonumber\\
c'' &=& -i\left[{2\omega_{p_t}(\omega_{p_t}-m_1-E_{(3/2,l=1)})\over
E_{(3/2,l=1)}-p_l)+i\epsilon}{2m+p_l\over (p_l+m_1)^2-\omega^{2}_{p_t}+i\epsilon}
\right]\nonumber\\
&&\times (2m_2(p_l-E_{(3/2,l=1)})\tilde\Phi^{(31)}_2;\nonumber\\
d'' &=& -i\left[{2\omega_{p_t}(\omega_{p_t}-m_1-E_{(3/2,l=1)})\over
E_{(3/2,l=1)}-p_l)+i\epsilon}{1\over (p_l+m_1)^2-\omega^{2}_{p_t}+i\epsilon}
\right] \cdot \nonumber\\
&& \times (2m_2(p_l-E_{(3/2,l=1)}))\tilde\Phi^{(31)}_2,
\end{eqnarray}
where $\tilde{\Phi}^{(s,l)}_i$ are the B-S wavefunctions after integrated
over $p_l$,
$$\tilde{\Phi}^{(s,l)}_i\equiv \int {dp_l\over 2\pi}\phi^{(s,l)}_i (p_l,
p_t^2),$$
$\omega_{p_t}=\sqrt{|p_t|^2+m_1^2}$, and we
have defined
$$\lambda_2={m_2\over m_1+m_2},$$
with $m_1$ being the light quark mass and $m_2$ the heavy diquark mass
($m_1\ll m_2$). $E_{(1/2, l)}$ and $E_{(3/2, l)}$ are binding energies
in the corresponding baryons.
%of $\Xi_{(bc)_0}$ and $\Xi^{**}_{(bc)}(s,l\geq 1)$ respectively.

All the functions are obtained  by carrying out the B-S integrations
which are very tedious, but straightforward (see ref.\cite{Guo}).

\end{document}